\documentclass[12pt,twoside]{article}

\usepackage{wrapfig}
\usepackage{graphicx}
\usepackage{cmp2e}

\title[A class of solvable models in CMP]%
{A class of solvable models \\ in Condensed Matter Physics}

\author{F. Mancini}

\address{Dipartimento di Fisica ''E.R. Caianiello" \\ Laboratorio
Regionale SuperMat, INFM \\ Universit\`{a} degli Studi di Salerno,
I-84081 Baronissi (SA), Italy}

\begin{document}

\maketitle

\begin{abstract}
In this paper, we show that there is a large class of fermionic
systems for which it is possible to find, for any dimension, a
finite closed set of eigenoperators and eigenvalues of the
Hamiltonian. Then, the hierarchy of the equations of motion closes
and analytical expressions for the Green's functions are obtained in
terms of a finite number of parameters, to be self-consistently
determined. Several examples are given. In particular, for these
examples it is shown that in the one-dimensional case it is possible
to derive by means of algebraic constraints a set of equations which
allow us to determine the self-consistent parameters and to obtain a
complete exact solution.


\pacs 71.10.-w; 71.10.Fd; 71.27.+a
\end{abstract}

One of the most intriguing problem in Condensed Matter Physics is
the study of highly interacting fermionic systems \cite{Fulde_95}.
The standard methods based on perturbation theories do not work well
and the attention is put on developing alternative approximate
formulations. In this context, it is interesting to cultivate the
study of some solvable models, which are themselves of physical
interest and which may furnish some indication on the solution of
more complex models. In a recent paper \cite{Mancini_05}, we have
shown that there is a large class of fermionic systems for which it
is possible to find a complete set of eigenoperators and eigenvalues
of the Hamiltonian. Then, the hierarchy of the equations of motion
closes and analytical exact expressions for the Green's functions
(GF) can be obtained.

We consider a system of $q$ species of particles, satisfying Fermi
statistics and localized on the sites of a Bravais lattice. We
suppose that the mass of the particles is very large and/or the
interaction is so strong that the kinetic energy is negligible and
the particles are frozen on the lattice sites. Let $c_{a}(i)$ and
$c_{a}^{\dag }(i)$ be the annihilation and creation operators of the
particles of species $a$ in the Heisenberg picture:
$i=(\mathbf{i},t)$, where $\mathbf{i}$ stands for the lattice vector
$R_{\mathbf{i}}$. These operators satisfy canonical anti-commutation
relations
\begin{equation}
\begin{array}{l}
\{c_{a}(\mathbf{i},t),c_{b}^{\dag }(\mathbf{j},t)\}=\delta
_{ab}\delta _{
\mathbf{ij}} \\
\{c_{a}(\mathbf{i},t),c_{b}(\mathbf{j},t)\}=\{c_{a}^{\dag
}(\mathbf{i} ,t),c_{b}^{\dag }(\mathbf{j},t)\}=0
\end{array}
\label{1}
\end{equation}
A quite general Hamiltonian for these systems reads as
\begin{equation}
H=\sum_{\mathbf{i}a}V_{a}(\mathbf{i})n_{a}(i)+\frac{1}{2}\sum_{\mathbf{ij}
}\sum_{ab}V_{ab}(\mathbf{i,j})n_{a}(i)n_{b}(j)+\frac{1}{3!}\sum_{\mathbf{ijl}
}\sum_{abc}V_{abc}(\mathbf{i,j,l})n_{a}(i)n_{b}(j)n_{c}(l)+\cdots
\label{2}
\end{equation}
where $V_{a}(\mathbf{i})$ represents an external field acting on the
particle $a$; $n_{a}(i)=c_{a}^{\dag }(i)c_{a}(i)$ is the particle
density operator of the species $a$; $V_{ab}$, $V_{abc}$ , $\cdots$
are the two-body, three-body, $\ldots$ potentials. In general, the
potentials will be translationally invariant and will depend only on
the differences between the coordinates of the particles.

The class of systems described by the Hamiltonian (\ref{2}) is very
large and of direct physical interest. By appropriate choice of the
potentials it may describe multiband electronic systems dominated by
charge correlations and by magnetic interactions in the
$s_{z}$-channel, where $s_{z}$ is the third component of the spin.
Furthermore, let $n(i)=\sum_{a}c_{a}^{\dag }(i)c_{a}(i)$ be the
total particle density and consider the transformation
$n(i)=\frac{q}{2}+S(i)$. Under this transformation the Hamiltonian
(\ref{2}) is mapped to spin-$q/2$ Ising-like models.

Now, if the interaction potentials have a finite range, the model
Hamiltonian is always solvable. The proof of this statement is the
following. Firstly, it is immediate to see that the density operator
$n_{a}(i)$ does not depend on time
\begin{equation}
\mathrm{i}\frac{\partial }{\partial t}n_{a}(i)=[n_{a}(i),H]=0
\label{3}
\end{equation}
Then, in order to use the equations of motion and GF formalism we
must start from the Heisenberg equation for the fermionic field. It
is immediate to see that $c_{a}(i)$ satisfies the equation of motion
\begin{equation}
\mathrm{i}\frac{\partial }{\partial
t}c_{a}(i)=V_{a}(\mathbf{i})c_{a}(i)+\sum_{
\mathbf{j}b}V_{ab}(\mathbf{i,j})n_{b}(j)c_{a}(i)+\frac{1}{2}\sum_{\mathbf{jl}
}\sum_{bc}V_{abc}(\mathbf{i,j,l})n_{b}(j)n_{c}(l)c_{a}(i)+\cdots
\label{4}
\end{equation}

The dynamics has generated other field operators of higher
complexity. By taking time derivatives of increasing order, more and
more complex operators are generated. These operators might be named
composite operators, as they are all expressed in terms of the
original fields $c_{a}(i)$. Because of the finite range of the
interaction, the higher-order field operators generated by the
dynamics extend over a finite number of sites. Then, because the
particle operator satisfies the algebra $[n_{a}(i)]^{p}=n_{a}(i)$
(for $ p\geq 1$), the number of composite operators is finite and a
complete set of eigenoperators of the Hamiltonian can be found. If
$n$ is the number of independent composite operators, we can
construct a $n$-multiplet operator
\begin{equation}
\psi (i)=\left(
\begin{array}{c}
\psi _{1}(i) \\
\psi _{2}(i) \\
\vdots \\
\psi _{n}(i)
\end{array}
\right)  \label{5}
\end{equation}
which satisfies the Heisenberg equation
\begin{equation}
\mathrm{i}\frac{\partial }{\partial t}\psi (i)=[\psi (i),H]=\epsilon
\psi (i) \label{6}
\end{equation}
where the $n\times n$ matrix $\epsilon $ will be denominated as the
energy matrix. Note that we are using a vectorial notation: the
field $\psi_{m}(i)$ is itself a multiplet of rank $q$. Once the
composite operator and the energy matrix have been determined, an
exact solution of the Hamiltonian can be formally obtained. Let us
define the retarded Green's function
\begin{equation}
G(i,j)=\left\langle R[\psi (i)\psi ^{\dag }(j)]\right\rangle =\theta
(t_{i}-t_{j})\left\langle \{\psi (i),\psi ^{\dag }(j)\}\right\rangle
\label{7}
\end{equation}
where $\left\langle \cdots \right\rangle $ denotes the
quantum-statistical average over the grand canonical ensemble. By
means of the Heisenberg equation (\ref{6}), we obtain in momentum
space the equation
\begin{equation}
[ \omega -\epsilon ]G(\mathbf{k},\omega )=I(\mathbf{k}) \label{8}
\end{equation}
where $I(\mathbf{k})$ is the Fourier transform of the normalization
matrix, defined as
\begin{equation}
I(\mathbf{i,j})=\left\langle \{\psi (\mathbf{i},t),\psi ^{\dag
}(\mathbf{j} ,t)\}\right\rangle  \label{9}
\end{equation}
The solution of Eq.~(\ref{8}) is
\begin{equation}
G(\mathbf{k},\omega )=\sum_{m=1}^{n}\frac{\sigma
^{(m)}(\mathbf{k})}{\omega -E_{m}+\mathrm{i}\delta }  \label{10}
\end{equation}
where $E_{m}$ are the eigenvalues of the energy matrix $\epsilon $.
The spectral density matrices $\sigma ^{(m)}(\mathbf{k})$ are
calculated by means of the formula \cite{Mancini_00}
\begin{equation}
\sigma _{\alpha \beta }^{(m)}(\mathbf{k})=\Omega _{\alpha
m}\sum_{\gamma }[\Omega _{m\gamma }]^{-1}I_{\gamma \beta
}(\mathbf{k})  \label{11}
\end{equation}
where $\Omega $ is the matrix whose columns are the eigenvectors of
the matrix $\epsilon $. The correlation function
$C(i,j)=\left\langle \psi (i)\psi ^{\dag }(j)\right\rangle $ can be
immediately calculated from (\ref{10}) and one obtains
\begin{equation}
\begin{array}{l}
C(i,j)=\frac{1}{N}\sum_{\mathbf{k}}\frac{1}{2\pi }\int d\omega
e^{i\mathbf{k} \cdot
(\mathbf{R}_{i}-\mathbf{R}_{j})-iE_{m}(t_{i}-t_{j})}C(\mathbf{k}
,\omega ) \\
C(\mathbf{k},\omega )=\pi \sum_{m=1}^{n}\delta (\omega
-E_{m})T_{m}\sigma ^{(m)}(\mathbf{k})
\end{array}
\label{12}
\end{equation}
with $T_{m}=1+\tanh \left( \beta E_{m}/2\right) $, $\beta $ is the
inverse temperature. By similar technique we can easily calculate
\cite{Mancini_05b} multi-point correlation functions as
$C(i,j;l_{1},l_{2},\cdots l_{s})=\left\langle \psi (i)\psi ^{\dag
}(j)n(l_{1})n(l_{2})\cdots n(l_{s})\right\rangle $.

Equations~(\ref{10}) and (\ref{12}) are an exact solution of the
model Hamiltonian (\ref{2}). However, the knowledge of the GF is not
fully achieved yet. The algebra of the field $\psi (i)$ is not
canonical: as a consequence, the normalization matrix
$I(\mathbf{k})$ in the equation of motion (\ref{8}) contains some
unknown static correlation functions. Generally, these correlators
are expectation values of operators not belonging to the chosen
basis and should be self-consistently calculated. According to the
scheme of calculations proposed by the composite operator method
\cite{Mancini_00,Mancini_04}, one way of calculating the unknown
correlators is by specifying the representation where the GF are
realized. The knowledge of the Hamiltonian and of the operatorial
algebra is not sufficient to completely determine the GF. The GF
refer to a specific representation (i.e., to a specific choice of
the Hilbert space) and this information must be supplied to the
equations of motion that alone are not sufficient to completely
determine the GF. The procedure is the following. From the algebra
it is possible to derive several relations among the operators. We
will call algebra constraints (AC) all possible relations among the
operators dictated by the algebra. This set of relations valid at
microscopic level must be satisfied also at macroscopic level, when
expectation values are considered. Use of these considerations leads
to some self-consistent equations which will be used to fix the
unknown correlators appearing in the normalization matrix. An
immediate set of rules is given by the equation
\begin{equation}
\left\langle \psi (i)\psi ^{\dag }(i)\right\rangle
=\frac{1}{N}\sum_{\mathbf{ k}}\frac{1}{2\pi }\int d\omega
C(\mathbf{k},\omega )  \label{13}
\end{equation}
where the l.h.s. is fixed by the AC and the boundary conditions
compatible with the phase under investigation, while in the r.h.s.
the correlation function $C(\mathbf{k},\omega )$ is computed by
means of the expression (\ref{12}). Use of (\ref{13}) will lead to a
set of exact relations among the correlation functions.
Unfortunately, for the considered class of systems, the number of
equations is always less than the number of correlation functions
and extra equations are needed. To this purpose we have recently
found \cite{Mancini_05,Mancini_05b} a set of AC capable to give
other relations among the correlation functions. Let us suppose that
there exist some operators, $O$, which project out of the
Hamiltonian a reduced part
\begin{equation}
OH=OH_{0}  \label{14}
\end{equation}
When $H_{0}$ and $H_{I}=H-H_{0}$ commute, an important rule which
descends from the AC (\ref{14}) is that the quantum statistical
averages over the complete Hamiltonian $H$ must coincide with the
average over the reduced Hamiltonian $H_{0}$
\begin{equation}
Tr\{Oe^{-\beta H}\}=Tr\{Oe^{-\beta H_{0}}\}  \label{15}
\end{equation}
For a variety of one-dimensional models, we have found that use of
the condition (\ref{15}) allows us to close the set of equations for
the correlation functions and to obtain an exact solution of the
model. For dimensions higher than one these conditions are not
sufficient and more equations are necessary.

We now apply the above procedure to some specific models, obtained
by special choice of the interaction potentials. We shall consider a
$d$-dimensional hypercubic Bravais lattice.

\textbf{First example}

We consider a $q$-state model described by the Hamiltonian
\begin{equation}
H=-\mu \sum_{\mathbf{i}}n(i)+dV\sum_{\mathbf{i}}n(i)n^{\alpha }(i)
\label{16}
\end{equation}
where $\mu $ is the chemical potential, $n(i)=\sum_{a}n_{a}(i)$ is
the total particle density. Hereafter, for a generic operator
$\Phi(i)$ we us the notation $\Phi ^{\alpha
}(i)=\sum_{\mathbf{j}}\alpha _{\mathbf{ij}}\Phi ( \mathbf{j},t)$,
where $\alpha _{\mathbf{ij}}$ is the projector on the first-nearest
neighbor sites. This model might describe a system of $q$-electron
bands, interacting through an intersite Coulomb potential of
strength $V$. Alternatively, by means of the transformation
$n(i)=\frac{q}{2} +S(i)$ the Hamiltonian (\ref{16}) takes the form
\begin{equation}
H=-h\sum_{\mathbf{i}}S(i)-J\sum_{\mathbf{i}}S(i)S^{\alpha }(i)+E_{0}
\label{17}
\end{equation}
where $J=-dV$, $h=\mu -qdV$, $E_{0}=\frac{q}{2}(-\mu
+\frac{q}{2}dV)N$; $N$ is the number of sites. Hamiltonian
(\ref{17}) is just the spin-$q/2$ Ising model with nearest neighbor
interactions in presence of an external magnetic field $h$. The
following recursion rule can be established \cite{Mancini_05b} for
the operator $[n^{\alpha }(i)]^{p}$
\begin{equation}
[  n^{\alpha }(i)]^{p}=\sum_{m=1}^{2qd}A_{m}^{(p)}[n^{\alpha
}(i)]^{m} \label{18}
\end{equation}
where the coefficients $A_{m}^{(p)}$ are rational numbers,
satisfying the relation $\sum_{m=1}^{2qd}A_{m}^{(p)}=1$, that can be
easily determined by the algebra and by the structure of the lattice
\cite{Mancini_05b}. Owing to this rule, the set of eigenoperators of
the Hamiltonian is given by
\begin{equation}
\psi (i)=\left(
\begin{array}{c}
c(i) \\
c(i)n^{\alpha }(i) \\
\vdots \\
c(i)[n^{\alpha }(i)]^{2qd}
\end{array}
\right) \qquad \qquad \qquad [\psi (i),H]=\epsilon \psi (i)
\label{19}
\end{equation}
where the energy matrix $\epsilon $ can be calculated by means of
the equation of motion and the recursion rule (\ref{18}). The
eigenvalues $E_{m}$ of $\epsilon $ are given by
\begin{equation}
E_{m}=-\mu +(m-1)V\qquad \qquad (m=1,2,\cdots 2qd+1)  \label{20}
\end{equation}
The retarded Green's functions and the correlation functions can be
exactly calculated by applying the scheme of calculations
illustrated above. Then, by using the AC (\ref{15}) it is possible
to derive the self-consistent equations
\begin{equation}
\kappa ^{(k-1)}-\lambda
^{(k-1)}=\frac{1}{2}\sum_{m=1}^{2qd+1}T_{m}\sigma _{1,k}^{(m)}\qquad
\qquad (k=1,\cdots, 2qd+1)  \label{21}
\end{equation}
where $T_{m}=1+\tanh \left( \beta E_{m}/2\right) $ and $\sigma
^{(m)}$ are the spectral density matrices, that can be calculated by
means of (\ref{11}) and are expressed in terms of the quantities
$\kappa ^{(p)}$. Equations (\ref{21}) give a set of exact relations,
valid for any value of $q$ and $d$, among the correlation functions
\begin{equation}
\begin{array}{l}
\kappa ^{(p)}=\left\langle [n^{\alpha }(i)]^{p}\right\rangle \\
\lambda ^{(p)}=\left\langle n(i)[n^{\alpha }(i)]^{p}\right\rangle
\end{array}
\qquad \qquad (p=0,1,\cdots 2qd)  \label{22}
\end{equation}
Unfortunately, the number of equations is not sufficient and we need
other conditions. We have studied the cases of $q=1$ ,$2$, $3$ in
Refs.~\cite{Mancini_05b,Mancini_05a,Avella_05}, respectively. In all
these cases for one-dimensional systems, it is possible to find by
means of the algebraic constraints (\ref{15}) the necessary
equations to close the set. These extra conditions are obtained by
exploiting the following algebraic conditions. Let us divide the
Hamiltonian as
\begin{equation}
H=H_{0}+2Vn(i)n^{\alpha }(i)  \label{23}
\end{equation}
where, due to translational invariance, $\mathbf{i}$ is a generic
site of the infinite chain. Then, we have
\begin{equation}
q=1\qquad c^{\dag }(i)e^{-\beta H}=c^{\dag }(i)e^{-\beta H_{0}}
\label{24}
\end{equation}
\begin{equation}
q=2\qquad
\begin{array}{l}
\xi ^{\dag }(i)e^{-\beta H}=\xi ^{\dag }(i)e^{-\beta H_{0}} \\
\eta ^{\dag }(i)e^{-\beta H}=\eta ^{\dag
}(i)\{1+{\sum\limits_{m=1}^{4}} f_{m}[n^{\alpha }(i)]^{m}\}e^{-\beta
H_{0}}
\end{array}
\label{25}
\end{equation}
\begin{equation}
q=3\qquad
\begin{array}{l}
\begin{array}{l}
\xi ^{\dag }(i)e^{-\beta H}=\xi ^{\dag }(i)e^{-\beta H_{0}} \\
{\eta ^{\dag }(i)e^{-\beta H}=\eta ^{\dag
}(i)\{1+\sum\limits_{m=1}^{6}{}f_{m}[n^{\alpha }(i)]^{m}\}e^{-\beta
H_{0}}}
\end{array}
\\
{\zeta ^{\dag }(i)e^{-\beta H}=\zeta ^{\dag
}(i)\{1+\sum\limits_{m=1}^{6}{}(2f_{m}+g_{m})[n^{\alpha
}(i)]^{m}\}e^{-\beta H_{0}}}
\end{array}
\label{26}
\end{equation}
For $q=2$ the definitions are:
\begin{equation}
\begin{array}{l}
\xi _{a}(i)=[1-n(i)]c_{a}(i) \\
\eta _{a}(i)=n(i)c_{a}(i)
\end{array}
\label{27}
\end{equation}
The operators $\xi _{a}(i)$ and $\eta _{a}(i)$ induce the
transitions $ 0\Leftrightarrow 1$, $1\Leftrightarrow 2$,
respectively. For $q=3$ the definitions are:
\begin{equation}
\begin{array}{l}
{\xi _{a}(i)=[1-n(i)+D(i)]c_{a}(i)} \\
{\eta _{a}(i)=[n(i)-2D(i)]c_{a}(i)} \\
{\zeta _{a}(i)=D(i)c_{a}(i)}
\end{array}
\label{28}
\end{equation}
where $D(i)$ is the double occupancy operator, defined as
\begin{equation}
{D(i)=n_{1}(i)n_{2}(i)+n_{1}(i)n_{3}(i)+n_{2}(i)n_{3}(i)}
\label{29}
\end{equation}
The projection operators $\xi _{a}$, $\eta _{a}$ and $\zeta _{a}$
induce the transitions among states with different particle numbers:
$0\Leftrightarrow 1$, $1\Leftrightarrow 2$, $2\Leftrightarrow 3$,
respectively. The quantities $f_{m}$ and $g_{m}$ are known functions
of $\beta V$.

\textbf{Second example}

Let us consider two species of particles, say $a$ and $b$, and
consider the Hamiltonian
\begin{equation}
H=-\mu
\sum_{\mathbf{i}}n(i)+U\sum_{\mathbf{i}}D(i)+dV\sum_{\mathbf{i}
}n(i)n^{\alpha }(i)  \label{30}
\end{equation}
where $n(i)=n_{a}(i)+n_{b}(i)$ and $D(i)=n_{a}(i)n_{b}(i)$ are the
total particle density and double occupancy operators, respectively.
This Hamiltonian is just the extended Hubbard model in the ionic
limit, where $U$ and $V$ are the on-site and inter-site Coulomb
interaction, respectively. The two species of particles, $a$ and
$b$, are in this case electrons with spin up and down, respectively.
By means of the transformation $n(i)=1+S(i)$ , (\ref{30}) can be
cast in the form
\begin{equation}
H=-h\sum_{\mathbf{i}}S(i)+\Delta
\sum_{\mathbf{i}}S^{2}(i)-J\sum_{\mathbf{i} }S(i)S^{\alpha
}(i)+E_{0}  \label{31}
\end{equation}
where $J=-dV$, $h=\mu -2dV-\frac{1}{2}U$, $\Delta =\frac{1}{2}U$, $
E_{0}=(-\mu +dV)N$. Hamiltonian (\ref{31}) is just the Ising
spin-$1$ model with nearest-neighbor interactions in the presence of
a crystal field $\Delta$ and an external magnetic field $h$
\cite{Blume_66,Capel_66,Capel_67,Capel_67a}. We now define the
composite operators
\begin{equation}
\psi ^{(\xi )}(i)=\left(
\begin{array}{c}
\xi (i) \\
\xi (i)n^{\alpha }(i) \\
\vdots \\
\xi (i)[n^{\alpha }(i)]^{4d}
\end{array}
\right) \qquad \qquad \psi ^{(\eta )}(i)=\left(
\begin{array}{c}
\eta (i) \\
\eta (i)n^{\alpha }(i) \\
\vdots \\
\eta (i)[n^{\alpha }(i)]^{4d}
\end{array}
\right)  \label{32}
\end{equation}
where $\xi (i)$ and $\eta (i)$ are the Hubbard operators, defined in
(\ref{27}). By means of (\ref{18}), these fields are eigenoperators
of the Hamiltonian (\ref{30})
\begin{equation}
\begin{array}{l}
\mathrm{i}\frac{\partial }{\partial t}\psi ^{(\xi )}(i)=[\psi ^{(\xi
)}(i),H]=\epsilon ^{(\xi )}\psi ^{(\xi )}(i) \\
\mathrm{i}\frac{\partial }{\partial t}\psi ^{(\eta )}(i)=[\psi
^{(\eta )}(i),H]=\epsilon ^{(\eta )}\psi ^{(\eta )}(i)
\end{array}
\label{33}
\end{equation}
where $\epsilon ^{(\xi )}$ and $\epsilon ^{(\eta )}$ are the energy
matrices, of rank $(4d+1)\times (4d+1)$ , which can be calculated by
means of the equations of motion and the recursion rule (\ref{18}).
The eigenvalues and of the energy matrices are given by
\begin{equation}
\begin{array}{l}
E_{m}^{(\xi )}=-\mu +(m-1)V\qquad \qquad \\
E_{m}^{(\eta )}=-\mu +U+(m-1)V
\end{array}
\qquad (m=1,2,\cdots 4qd+1)  \label{34}
\end{equation}
The retarded and the correlation functions can be exactly calculated
by applying the scheme of calculations illustrated above. Then, by
using the AC (\ref{13}) it is possible to derive the self-consistent
equations
\begin{equation}
\kappa ^{(k-1)}-\frac{1}{2}\lambda ^{(k-1)}=\frac{1}{2}
\sum_{m=1}^{4d+1}[T_{m}^{(\xi )}\sigma _{1,k}^{(\xi
,m)}+T_{m}^{(\eta )}\sigma _{1,k}^{(\eta ,m)}]\qquad \qquad
(k=1,\cdots 4d+1)  \label{35}
\end{equation}
where $T_{m}^{(a)}=1+\tanh \left( \beta E_{m}^{(a)}/2\right) $ with
$a=\xi ,\eta $. $\sigma ^{(a,m)}$ are the spectral density matrices,
that can be calculated by means of (\ref{11}) and are expressed in
terms of the quantities $\kappa ^{(p)}$ and $\lambda ^{(p)}$.
Equations (\ref{35}) give a set of exact relations, valid for any
dimension, among the correlation functions (\ref{22}).
Unfortunately, the number of equations is not sufficient and we need
other conditions. Our study \cite{Mancini_05a} has shown that for
the one-dimensional case, it is possible to find by means of the
algebraic constraints (\ref{15}) the necessary equations to close
the set (\ref{35}). These extra conditions are obtained by
exploiting the following algebraic conditions
\begin{equation}
\begin{array}{l}
\xi ^{\dag }(i)e^{-\beta H}=\xi ^{\dag }(i)e^{-\beta H_{0}} \\
D(i)e^{-\beta H}= D(i)\{1+{\sum\limits_{m=1}^{4}(2}
f_{m}+g_{m})[n^{\alpha }(i)]^{m}\}e^{-\beta H_{0}}
\end{array}
\label{36}
\end{equation}
where $H_{0}=H-2Vn(i)n^{\alpha }(i)$.

\textbf{Third example}

The model (\ref{30}) can be generalized by considering 3- and 4-body
potentials. In particular, by considering one-dimensional systems,
let us take
\begin{equation}
H=-\mu \sum_{\mathbf{i}}n(i)-\gamma
\sum_{\mathbf{i}}D(i)+V\sum_{\mathbf{i} }n(i)n^{\alpha
}(i)+U\sum_{\mathbf{i}}D(i)D^{\alpha }(i)-U\sum_{\mathbf{i}
}D(i)n^{\alpha }(i)  \label{37}
\end{equation}

This Hamiltonian can be mapped to the following model
\begin{equation}
H=-h\sum_{\mathbf{i}}S(i)+\Delta
\sum_{\mathbf{i}}S^{2}(i)-J\sum_{\mathbf{i} }S(i)S^{\alpha
}(i)-K\sum_{\mathbf{i}}S^{2}(i)S^{2\alpha }(i)+E_{0} \label{38}
\end{equation}
where $J=\frac{1}{4}U-V$, $K=-\frac{1}{4}U$, $h=\mu
+\frac{1}{2}\gamma + \frac{1}{2}U-2V$, $\Delta
=-\frac{1}{2}U-\frac{1}{2}\gamma $, $E_{0}=(-\mu +V)N$. This
Hamiltonian is just the one-dimensional Blume-Emery-Griffiths model
\cite{Blume_71}. We now define the composite operators
\begin{equation}
\psi ^{(\xi )}(i)=\left(
\begin{array}{c}
\xi (i) \\
\xi (i)n^{\alpha }(i) \\
\xi (i)[n^{\alpha }(i)]^{2} \\
\xi (i)[n^{\alpha }(i)]^{3} \\
\xi (i)[n^{\alpha }(i)]^{4} \\
\xi (i)D^{\alpha }(i) \\
\xi (i)[D^{\alpha }(i)]^{2}
\end{array}
\right) \qquad \qquad \psi ^{(\eta )}(i)=\left(
\begin{array}{c}
\eta (i) \\
\eta (i)n^{\alpha }(i) \\
\eta (i)[n^{\alpha }(i)]^{2} \\
\eta (i)[n^{\alpha }(i)]^{3} \\
\eta (i)[n^{\alpha }(i)]^{4} \\
\eta (i)D^{\alpha }(i) \\
\eta (i)[D^{\alpha }(i)]^{2}
\end{array}
\right)  \label{39}
\end{equation}
where $\xi (i)$ and $\eta (i)$ are the Hubbard operators, [cfr.
(\ref{27})]. These fields are eigenoperators of the Hamiltonian
(\ref{37}). The corresponding eigenvalues are
\begin{equation}
E_{m}^{(\xi )}=\left(
\begin{array}{c}
-\mu \\
-\mu +V \\
-\mu +2V \\
-\mu -U+2V \\
-\mu -\frac{1}{2}U+2V \\
-\mu -U+4V \\
-\mu -\frac{1}{2}U+3V
\end{array}
\right) \qquad \qquad E_{m}^{(\eta )}=\left(
\begin{array}{c}
-\mu -\gamma \\
-\mu -\gamma -\frac{1}{2}U+V \\
-\mu -\gamma -U+2V \\
-\mu -\gamma +2V \\
-\mu -\gamma -\frac{1}{2}U+2V \\
-\mu -\gamma -U+4V \\
-\mu -\gamma -U+3V
\end{array}
\right)  \label{40}
\end{equation}
The retarded and the correlation functions can be exactly calculated
by applying the scheme of calculations illustrated above. Then, by
using the AC (\ref{13}) it is possible to derive \cite{Mancini_05c}
the self-consistent equations
\begin{equation}
\begin{array}{l}
\lambda ^{(0)}-\delta ^{(1)}=\frac{1}{2}\sum_{m=1}^{7}T_{m}^{(\eta
)}\sigma
_{1,1}^{(\eta ,m)} \\
\kappa ^{(k-1)}-\frac{1}{2}\lambda ^{(k-1)}=\frac{1}{2}
\sum_{m=1}^{7}[T_{m}^{(\xi )}\sigma _{1,k}^{(\xi ,m)}+T_{m}^{(\eta
)}\sigma
_{1,k}^{(\eta ,m)}]\qquad \qquad (k=1,\cdots 5) \\
\delta ^{(k-5)}-\frac{1}{2}\theta ^{(k-5)}=\frac{1}{2}
\sum_{m=1}^{7}[T_{m}^{(\xi )}\sigma _{1,k}^{(\xi ,m)}+T_{m}^{(\eta
)}\sigma _{1,k}^{(\eta ,m)}]\qquad \qquad (k=6,7)
\end{array}
\label{41}
\end{equation}
where the definitions are the same as in example 2, and the new
correlations functions and are defined as
\begin{equation}
\begin{array}{l}
\delta ^{(p)}=\left\langle [D^{\alpha }(i)]^{p}\right\rangle \\
\theta ^{(p)}=\left\langle n(i)[D^{\alpha }(i)]^{p}\right\rangle
\end{array}
\qquad \qquad (p=1,2)  \label{42}
\end{equation}
$\sigma ^{(a,m)}$ are the spectral density matrices and are
expressed in terms of the quantities $\kappa ^{(p)},\lambda
^{(p)},\delta ^{(p)},\theta ^{(p)}$. Equations (\ref{41}) give a set
of exact relations among the correlation functions (\ref{22}) and
(\ref{42}). By exploiting the algebraic condition $\xi ^{\dag
}(i)e^{-\beta H}=\xi ^{\dag }(i)e^{-\beta H_{0}}$ it is possible to
derive \cite{Mancini_05c} the self-consistent equations
\begin{equation}
\begin{array}{l}
C_{1,3}^{(\xi )}=C_{1,1}^{(\xi )}[\frac{1}{2}X_{1}+X_{2}+\frac{1}{2}
X_{1}^{2}] \\
C_{1,4}^{(\xi )}=C_{1,1}^{(\xi
)}[\frac{1}{4}X_{1}+\frac{3}{2}X_{2}+\frac{3}{
2}X_{1}X_{2}+\frac{3}{4}X_{1}^{2}] \\
C_{1,5}^{(\xi )}=C_{1,1}^{(\xi
)}[\frac{1}{8}X_{1}+\frac{7}{4}X_{2}+\frac{9}{
2}X_{1}X_{2}+\frac{7}{8}X_{1}^{2}+\frac{3}{2}X_{2}^{2}] \\
C_{1,7}^{(\xi )}=C_{1,1}^{(\xi
)}[\frac{1}{2}X_{2}+\frac{1}{2}X_{2}^{2}]
\end{array}
\label{43}
\end{equation}
where
\begin{equation}
X_{1}=\frac{C_{1,2}^{(\xi )}}{C_{1,1}^{(\xi )}}\qquad X_{2}=\frac{
C_{1,6}^{(\xi )}}{C_{1,1}^{(\xi )}}  \label{44}
\end{equation}
$C_{1,k}^{(\xi )}$ is the equal time correlation function
$C_{1,k}^{(\xi )}=\left\langle \psi _{1}^{(\xi )}(i)\psi _{1}^{(\xi
)\dag }(i)\right\rangle $, expressed in terms of the parameters
$\kappa ^{(p)},\lambda ^{(p)},\delta ^{(p)},\theta ^{(p)}$ by means
of the relation
\begin{equation}
C_{1,k}^{(\xi )}=\frac{1}{2}\sum_{m=1}^{7}T_{m}^{(\xi )}\sigma
_{1,k}^{(\xi ,m)}  \label{45}
\end{equation}

\textbf{Fourth example}

Let us consider the case of two particles, characterized by spin
$\sigma =\uparrow ,\downarrow (+,-)$, and choice the potentials as
\begin{equation}
\begin{array}{l}
V_{\sigma }(\mathbf{i})=-\mu -\sigma h \\
V_{\sigma ,\sigma ^{\prime }}(\mathbf{i,j})=2Vd(2\delta _{\sigma
,\sigma ^{\prime }}-1)\alpha _{\mathbf{i,j}}
\end{array}
\label{46}
\end{equation}
$h$ is the intensity of the external magnetic field. The Hamiltonian
is
\begin{equation}
H=-\mu
\sum_{\mathbf{i}}n(i)+dV\sum_{\mathbf{i}}n_{3}(i)n_{3}^{\alpha
}(i)-h\sum_{\mathbf{i}}n_{3}(i)  \label{47}
\end{equation}
where $n_{3}(i)=n_{\uparrow }(i)-n_{\downarrow }(i)$ is the third
component of the spin density operator. Let us restrict the analysis
to one-dimensional systems. The following recursion rule can be
established for the operator $[n_{3}^{\alpha }(i)]^{p}$
\begin{equation}
[  n_{3}^{\alpha }(i)]^{p}=\sum_{m=1}^{4}A_{m}^{(p)}[n_{3}^{\alpha
}(i)]^{m} \label{48}
\end{equation}
where the coefficients $A_{m}^{(p)}$ are rational numbers,
satisfying the relation $\sum_{m=1}^{4}A_{m}^{(p)}=1$, that can be
easily determined by the algebra and the structure of the lattice.
Owing to this rule, the set of eigenoperators of the Hamiltonian
(\ref{47}) is given by
\begin{equation}
\psi (i)=\left(
\begin{array}{l}
\psi _{\uparrow }(i) \\
\psi _{\downarrow }(i)
\end{array}
\right) \qquad \qquad \psi _{\sigma }(i)=\left(
\begin{array}{l}
c_{\sigma }(i) \\
c_{\sigma }(i)n_{3}^{\alpha }(i) \\
c_{\sigma }(i)[n_{3}^{\alpha }(i)]^{2} \\
c_{\sigma }(i)[n_{3}^{\alpha }(i)]^{3} \\
c_{\sigma }(i)[n_{3}^{\alpha }(i)]^{4}
\end{array}
\right)  \label{49}
\end{equation}
\begin{equation}
\mathrm{i}\frac{\partial }{\partial t}\psi _{\sigma }(i)=\epsilon
^{(\sigma )}\psi _{\sigma }(i)  \label{50}
\end{equation}
The energy matrix $\epsilon ^{(\sigma )}$ is a $5 \times 5$ matrix
with the following expression
\begin{equation}
\epsilon ^{(\sigma )}=\left(
\begin{array}{ccccc}
-\mu -\sigma h & 2\sigma V & 0 & 0 & 0 \\
0 & -\mu -\sigma h & 2\sigma V & 0 & 0 \\
0 & 0 & -\mu -\sigma h & 2\sigma V & 0 \\
0 & 0 & 0 & -\mu -\sigma h & 2\sigma V \\
0 & -\frac{1}{2}\sigma V & 0 & \frac{5}{2}\sigma V & -\mu -\sigma h
\end{array}
\right)  \label{51}
\end{equation}
The eigenvalues of the matrix $\epsilon ^{(\sigma )}$ are
\begin{equation}
E_{m}^{(\sigma )}=\left(
\begin{array}{c}
-\mu -\sigma h \\
-\mu -\sigma h-2V \\
-\mu -\sigma h-V \\
-\mu -\sigma h+V \\
-\mu -\sigma h+2V
\end{array}
\right)  \label{52}
\end{equation}
The retarded Green's functions and the correlation functions can be
exactly calculated by applying the scheme of calculations
illustrated above. Then, by using the AC (\ref{13}) it is possible
to derive the self-consistent equations
\begin{equation}
\begin{array}{l}
2\kappa ^{(k-1)}-\lambda
^{(k-1)}=\frac{1}{2}\sum_{m=1}^{5}[T_{m}^{(\uparrow )}\sigma
_{1,k}^{(\uparrow ,m)}+T_{m}^{(\downarrow )}\sigma
_{1,k}^{(\downarrow ,m)}] \\
\theta ^{(k-1)}=\frac{1}{2}\sum_{m=1}^{5}[T_{m}^{(\uparrow )}\sigma
_{1,k}^{(\uparrow ,m)}-T_{m}^{(\downarrow )}\sigma
_{1,k}^{(\downarrow ,m)}]
\end{array}
\qquad (k=1,\cdots 5)  \label{53}
\end{equation}
where $T_{m}^{(\sigma )}=1+\tanh \left( \beta E_{m}^{(\sigma
)}/2\right) $ and $\sigma ^{(\sigma ,m)}$ are the spectral density
matrices, that can be calculated by means of (\ref{11})and are
expressed in terms of the quantities $\kappa ^{(p)}$. Equations
(\ref{53}) give a set of exact relations among the correlation
functions
\begin{equation}
\begin{array}{l}
\kappa ^{(p)}=\left\langle [n_{3}^{\alpha }(i)]^{p}\right\rangle \\
\lambda ^{(p)}=\left\langle n(i)[n_{3}^{\alpha }(i)]^{p}\right\rangle \\
\theta ^{(p)}=\left\langle n_{3}(i)[n_{3}^{\alpha
}(i)]^{p}\right\rangle
\end{array}
\qquad \qquad (p=0,1,\cdots 4)  \label{54}
\end{equation}
Also for this model we can use the AC (\ref{15}) in order to derive
extra equations capable to close the set (\ref{53}) for the
correlation functions. Details will be presented elsewhere.

Summarizing, we have shown that a large class of systems of
localized particles, satisfying Fermi statistics and subject to
finite-range interactions, is always solvable, in the sense that a
complete finite set of eigenoperators and eigenvalues of the
Hamiltonian can be found. This knowledge allows us to derive
analytical expressions for the Green's functions and for the
correlation functions and a set of exact relations among the
correlation functions can be derived. As an illustration we have
considered several examples. In all the studied models we have shown
that for the case of one dimension it is possible to use algebraic
constraints which permit to close the set of self-consistent
equations and to obtain exact solutions. For higher dimensions more
self-consistent equations are needed. This problem is now under
investigation.

\label{last@page}

\begin{thebibliography}{13}
\bibitem{Fulde_95} Fulde~P., Electron Correlations in Molecules and Solids. Springer-Verlag, Berlin, 1995.
\bibitem{Mancini_05} Mancini~F., Europhys.~Lett., 2005, \textbf{70}, 485.
\bibitem{Mancini_00} Mancini~F. and Avella A., Eur.~Phys.~J.~B, 2003, \textbf{36}, 37.
\bibitem{Mancini_05b} Mancini~F., Eur.~Phys.~J.~B, 2005, \textbf{45}, 497.
\bibitem{Mancini_04} Mancini~F. and Avella A., Adv.~Phys., 2004, \textbf{53}, 537.
\bibitem{Mancini_05a} Mancini~F., The extended Hubbard model in the ionic limit, Preprint University of Salerno, 2005. Submitted to Eur.~Phys.~J.~B.
\bibitem{Avella_05} Avella~A. and Mancini~F., Exact solution of the one-dimensional spin-3/2 Ising model in magnetic field, Preprint University of Salerno, 2005.
\bibitem{Blume_66} Blume~M., Phys.~Rev., 1966, \textbf{141}, 517.
\bibitem{Capel_66} Capel~H., Physica, 1966, \textbf{32}, 966.
\bibitem{Capel_67} Capel~H., Physica, 1967, \textbf{33}, 295.
\bibitem{Capel_67a} Capel~H., Physica, 1967, \textbf{37}, 423.
\bibitem{Blume_71} Blume~M., Emery~V., Griffiths~R., Phys.~Rev.~A, 1971, \textbf{4}, 1071.
\bibitem{Mancini_05c} Mancini~F., Mancini~F.P., in preparation, 2005.
\end{thebibliography}
\end{document}